\documentclass[9pt,onecolumn,twoside]{article}

\usepackage{lipsum}
\usepackage{authblk}
\usepackage{fancyhdr}
\usepackage{amsmath}
\usepackage{xfrac}
\usepackage{algpseudocode} 
\usepackage{algorithm}
\usepackage[english,plain]{fancyref}
\usepackage[margin=1in]{geometry}

\newcommand{\INT}[2]{\int {#1}\,\text{d}{#2}}
\newcommand{\DIFF}[2]{\frac{\text{d}}{\text{d}{#2}}({#1})}

\title{Frequency-resolved optical gating of highly chirped ultrabroadband pulses}

\author[1,2,*]{Adam S Wyatt}
\author[1]{Pedro Oliveira}
\author[1]{Ian O Musgrave}

\affil[1]{Central Laser Facility, STFC Rutherford Appleton Laboratory, Harwell OX11 0QX, UK}
\affil[2]{Clarendon Laboratory, Department of Physics, University of Oxford, Oxford OX1 3PU, UK}

\affil[*]{Corresponding author: adam.wyatt@stfc.ac.uk}

\begin{document}

\maketitle
\thispagestyle{fancy}
\abstract{
This article describes a simplified reconstruction algorithm for frequency resolved optical gating (FROG) measurements of highly (monotonically) chirped pulses. The FROG traces are calculated using the stationary phase approximation, significantly reducing the data size and convergence time compared to discrete Fourier transform based methods. The method is verified via second harmonic generation FROG simulations and measurements of pulses centred at 800\,nm with a bandwidth of 55\,nm stretched to 32\,ps at the 1\% intensity level, corresponding to a time-bandwidth product of 830.
}

\section{Introduction}
\label{sec:INTRO}
Chirped pulses are used in a variety of applications, for example in [optical parametric] chirped pulse amplification ([OP]CPA)~\cite{ross1997prospects, cerullo2003ultrafast, butkus2004progress}, chirp-assisted sum frequency generation (CA-SFG)~\cite{osvay1999efficient}, dispersive Fourier transform spectroscopy (DFTS)~\cite{goda2013dispersive} and telecommunications~\cite{boivin1997206}. DFTS normally requires linear dispersion over the pulse bandwidth to prevent distorting the data, but this constraint can be relaxed provided the chirp is monotonic and known~\cite{mahjoubfar2015design}. 
Extremely large bandwidth OPCPA systems require complex stretcher/compressor configurations to compensate for the material dispersion and OPA phase, hence composite systems are typically employed utilizing a combination of prism, grating and bulk stretchers/compressors in conjunction with an adaptive element~\cite{mikhailova2011ultra}. The cost and complexity of compressing these pulses makes it desirable to have a means to measure the amplified chirped pulses at various locations along the laser chain in order to diagnose and optimize each element of the chain in situations where a compressed pulse is not available and cannot easily be generated.

In chirp-compensated (CC-) OPCPA, appropriate nonlinear chirp is applied to the pump and signal pulses in order to increase the gain bandwidth in a collinear geometry~\cite{tang2008optical,wyatt2015ultra}. This is achieved by setting the instantaneous frequencies at any group delay of the pump and signal to be those that phase match in a long crystal. Since the frequency-dependent group delay (GD) of the pump and seed need to be matched to better than 100\,fs over the whole duration of the pulse (typically >10\,ps), a coarse measurement of the pulse duration is not sufficient and accurate measurement of the nonlinear dispersion is required.

The characterization of ultrashort laser pulses is a mature topic with a variety of methods available~\cite{walmsley2009characterization}. Although it is possible to measure the electric waveform of ultrashort pulses~\cite{frank2012invited, kim2013petahertz, wyatt2016attosecond}, these methods typically require near few-cycle waveforms and strong-field phenomena and so are not applicable to highly chirped pulses. More general methods include spectral phase interferometry for direct electric field reconstruction (SPIDER)~\cite{iaconis1998spectral}, frequency-resolved optical gating (FROG)~\cite{trebino1997measuring}, multiphoton intrapulse interference phase scan (MIIPS)~\cite{lozovoy2004multiphoton} and dispersion scan (DS)~\cite{miranda2012simultaneous}.

A limiting factor in SPIDER is the requirement to frequency mix with a pulse which is quasi-monochromatic over the duration of the test pulse. DS by design measures the pulse around zero net group delay dispersion (GDD), and MIPS requires the application of a phase function that can compensate the frequency dependent group delay across the whole spectrum. Experimental requirements for FROG include the ability to fully temporally scan the two time-delayed pulses through each other, and to fully resolve the FROG spectrum with sufficient signal to noise. The spectral phase and amplitude are then retrieved via an iterative algorithm. Typical algorithms are based on DFTs of the spectral/temporal amplitudes, for example the principle component generalized projection algorithm~\cite{kane2008principal}, which turn out to be unusable for highly chirped pulses. 

For any given pulse, the minimum number of points, $N$, required to sample the analytic field suitable for a DFT computation is simply given by the time-bandwidth product,
\begin{equation}
    \label{eq:TBP}
    N = \frac{BT}{2\pi},
\end{equation}
where $B$ and $T$ are the spectral and temporal support of the pulse at some given dynamic range. For SHG-FROG, the spectral support of the SHG spectrum will at most be twice that of the fundamental spectrum, i.e. $B_\text{SHG} = 2B_\text{Fund}$. The maximum temporal delay, $\tau$, between the two time-delayed replicas is simply be twice the temporal support of the pulse being measured, i.e. $\tau_\text{max}-\tau_\text{min} = 2T_\text{Fund}$. Therefore the number of points required to sample a SHG-FROG trace in which the temporal and spectral domains are related by a DFT is
\begin{equation}
    \label{eq:SHG_FROG_TBP}
    N_\text{SHG-FROG} = \left(\frac{2B_\text{Fund}\times2T_\text{Fund}}{2pi}\right)^2 = (4N_\text{Fund})^2.
\end{equation}

Taking $N_\text{Fund} = 2^{10} = 1024$ as an example would require $N_\text{SHG-FROG} = 2^{24} = 16,777,216$ points corresponding to 0.25\,GiB of data (double-precision, complex-valued). Whilst this information can easily be handled by modern computers, inclusion into even the most highly optimized DFT-based FROG algorithm will cause it to grind to a halt. Note that due to limited signal to noise and dynamic range, the measurable FROG signal above the noise floor will actually be contained within a smaller boundary than as just described, typically by a factor of 2--3 in each dimension. 

Another related issue is that there are, in this example, $2^{13}$ degrees of freedom for the algorithm to retrieve ($2^{12}$ degrees of freedom each for the spectral intensity and spectral phase). In practice, monotonically chirped pulses can typically be described by a low number of degrees of freedom ($N<100$) for the spectral intensity and a polynomial/smoothly varying spectral phase with a large average group delay dispersion. 
Not only will DFT algorithms take a long time per iteration, but they will require a vast number of iterations in order to sift through a significant number of mathematically possible but practically irrelevant solutions that have a degree of complexity much higher than that of the physical field being measured. In this article, I describe a new FROG algorithm that makes use of the stationary phase approximation (SPA) to significantly reduce the required number of sampling points, and hence computational complexity and retrieval time, when applied to the retrieval of the group delay dispersion of monotonically chirped ultrabroadband optical pulses.

\section{Algorithm}
\label{sec:ALGORITHM}

A monotonically chirped pulse is defined as one in which the group delay dispersion (GDD) is of constant sign and large magnitude relative to the bandwidth of the finest spectral feature, $\Delta\omega_\text{min}$: 
\begin{equation}
    \label{eq:SPA_CRITERION}
    \left|\phi^{\prime\prime}(\omega)\right| \gg 2\pi/\Delta\omega_\text{min}^2.
\end{equation}
Using the SPA, the measured FROG intensity can be written in terms of the spectral intensity and GDD of the test pulse. In the case of SHG-FROG, the measured trace can be written as
\begin{equation}
    \label{eq:SPA_FROG}
    F(\tau, \omega_3) \simeq \eta
    \left(\frac{I(\omega_1)}{n(\omega_1)\left|\phi^{\prime\prime}(\omega_1)\right|}\right)
    \left(\frac{I(\omega_2)}{n(\omega_2)\left|\phi^{\prime\prime}(\omega_2)\right|}\right)
    \left(\frac{\omega_3^2\left|\phi^{\prime\prime}(\omega_3)\right|}{n(\omega_3)}\right)
    \text{sinc}^2\left[\frac{\Delta k\left(\omega_1,\omega_2\right)L}{2}\right]
\end{equation}
where the constant $\eta$ relates to the SHG efficiency, $I(\omega) \propto n(\omega)|E(\omega)|^2$ is the spectral intensity,  
$\Delta k(\omega_1, \omega_2) = \left[n(\omega_3)\omega_3 - n(\omega_2)\omega_2 - n(\omega_1)\omega_1\right]/c$ is the wavevector mismatch, $c$ is the speed of light, $n(\omega)$ is the refractive index of the nonlinear medium, and
$\omega_3(\omega_1, \tau) = \omega_1 + \omega_2(\omega_1, \tau)$ is the sum frequency. Since the pulses are monotonically chirped, the frequency dependent group delay, $\phi^\prime(\omega)$, of any two interacting frequencies are constrained as
\begin{equation}
    \label{eq:FREQ_SHFT}
    \phi^{\prime}(\omega_1) = \phi^{\prime}(\omega_2) + \tau
\end{equation}

\Fref{eq:SPA_FROG} has a straightforward and intuitive explanation: the instantaneous sum frequency intensity is the product of the instantaneous intensities of the two mixing fields and a phase matching function. The large GDD results in a mapping of frequency to time such that the instantaneous electric field envelope
can be approximated using the SPA as 
\begin{equation}
    \label{eq:SPA}
    \left|E[t=\widetilde{\phi}^\prime(\omega)]\right|^2 \propto 
    \frac{I(\omega)}{n(\omega) \left|\phi^{\prime\prime}(\omega)\right|}.
    \end{equation}
The instantaneous sum-frequency generation (SFG) field envelope 
can be calculated directly using Maxwell's wave equation in the limit of no fundamental field depletion as 
\begin{equation}
    \label{eq:SFG}
    \left|E_\text{SFG}[t=\phi^\prime(\omega_3)]\right| \propto [k(\omega_3)/n(\omega_3)] E\left[\phi^\prime(\omega_1)\right] E\left[\phi(\omega_2)\right]
\end{equation}
and $\omega_3 = \omega_1 + \omega_2$.
The SFG signal is multiplied by a phase matching function, $\text{sinc}(\ldots)$, due to propagation through a crystal of finite thickness. The SFG spectral intensity is then the product of the refractive index and modulus square of the spectral field envelope after application of the SPA. A similar expression to \fref{eq:SPA_FROG} can be found for other nonlinearities. 
The procedure for numerically generating an SHG-FROG trace using is outlined in algorithm~\ref{alg:SPA_FROG}.

\begin{algorithm}
    \caption{Procedure for generating an SHG-FROG trace via the SPA.}
    \label{alg:SPA_FROG}
    \begin{algorithmic}[1]
        \Procedure{SHG\_FROG}{$\omega_1, \omega_3, \tau, I(\omega), \phi^{\prime\prime}(\omega), n(\omega), L$}
            \State $\text{GDw1} \gets \INT{\phi^{\prime\prime}(\omega_1)}{\omega_1}$
            %
            \State $\text{GDw2} \gets \text{GDw1} - \tau$
            %
            \State $\omega_2 \gets \text{interp}(\text{GDw1}, \omega_1, \text{GDw2})$
            %
            \State $\omega_3^\prime \gets \omega_1 + \omega_2$
            %
            \State $\text{GDDw1} \gets \phi^{\prime\prime}(\omega_1)$
            %
            \State $\text{GDDw2} \gets \phi^{\prime\prime}(\omega_2)$
            %
            \State $\text{GDDw3} \gets \DIFF{\text{GDw1}}{\omega_3^\prime}$
            \Comment{$\phi^\prime(\omega_1)=\phi^\prime(\omega_2)=\phi^\prime(\omega_3)$}
            %
            %
            %
            %
            \State $\text{Iw1} \gets I(\omega_1)/(n(\omega_1)\times\text{GDDw1})$
            %
            \State $\text{Iw2} \gets I(\omega_2)/(n(\omega_2)\times\text{GDDw2})$
            \State $\text{PMF} \gets \text{sinc}[(n(\omega_3)\omega_3 - n(\omega_2)\omega_2 - n(\omega_1)\omega_1)L/(2c)]$
            \State $\text{Fw3} \gets \text{Iw1}\times\text{Iw2} \times\text{GDDw3}\times{\omega_3^\prime}^2\times \text{PMF}/n(\omega_3^\prime)$
            %
            \State $F \gets \text{interp}(\omega_3^\prime, \text{Fw3}, \omega_3)$
            \State \textbf{return} $F$
        \EndProcedure
    \end{algorithmic}
\end{algorithm}

In order to reconstruct the test pulse spectral phase, the FROG error defined in \fref{eq:FROG_ERROR} is minimized via a direct search algorithm, e.g. downhill simplex with restarts.
\begin{equation}
    \label{eq:FROG_ERROR}
    \varepsilon = \sum_{i, j} 
    \left| \mu_m F_m(\tau_i, \omega_j) - \mu_r F_r(\tau_i, \omega_j) \right|/(N_iN_j)
\end{equation}
with  
$\mu_m = (N_iN_j)/\left[ \sum_{i, j} F(\tau_i, \omega_j) \right]$ 
used to normalize the the FROG trace to unity average intensity. The definition of the FROG error as defined in \fref{eq:FROG_ERROR} differs slightly to that typically found in the literature to enable an easy comparison for different sampling rates. The reconstructed FROG trace is normalized to minimizes the FROG error using\\
$\mu_r = \mu_m\left[\sum_{i,j} F_m(\tau_i,\omega_j) F_r(\tau_i,\omega_j)\right] / \sum_{i,j} F_r^2(\tau_i, \omega_j)$.
It is important that the reconstructed FROG trace is normalized by a parameter that depends globally on the measured FROG trace and not locally, otherwise noise at the locally measured location will effectively be added to every point across the measured trace, significantly distorting the reconstruction error.

\section{Simulations}
\label{sec:SIMULATIONS}

\begin{figure}[htbp]
\centering
\fbox{\includegraphics[width=\linewidth]{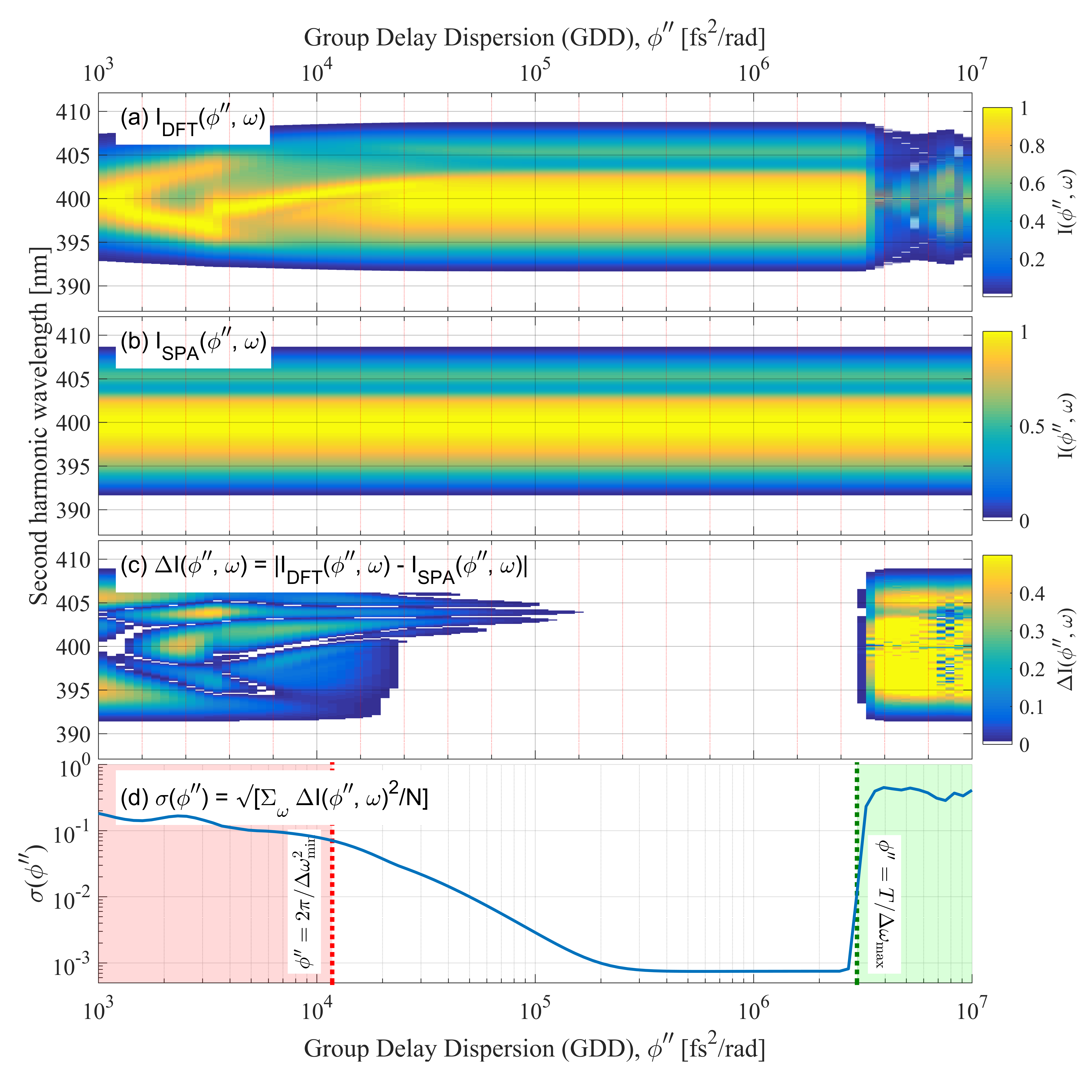}}
\caption{Accuracy of the SPA as a function of GDD determined by the SHG spectrum resulting from type I doubling in BBO. (a) Calculation using a DFT based nonlinear differential equation solver. (b) Calculation using the SPA. (c) Absolute difference between the DFT and SPA calculations. (d) RMS error as a function of GDD with the limits of validity marked by the dotted lines (see text for details).}
\label{fig:GDD_DEP}
\end{figure}

\Fref{fig:GDD_DEP} shows the accuracy of using the SPA to calcualte the second harmonic spectrum of a pulse as a function of group delay dispersion. The spectrum of the pulse was chosen to be a super Gaussian of bandwidth $\Delta\omega \sim 30$\,nm centered at 800\,nm with a $\Delta\omega_\text{min} \sim 3$\,nm spectral hole centred at 808\,nm and 30\% intensity. The corresponding Fourier transform limited (FTL) full width at half maximum (FWHM) pulse duration is $\Delta t \sim 55$\,fs. A linear chirp, quantified by $\phi^{\prime\prime}(\omega) = \phi^{\prime\prime}$ is applied to the spectrum and the type I second harmonic spectrum generated in BBO calculated via two methods. The first method evenly samples the spectral amplitude with $N=2^{14}$ points and bandwidth of $B \sim 0.3$\,rad/fs ($\Delta\lambda=100$\,nm) resulting in a temporal window of $T \sim 350$\,ps and then solves the first order nonlinear propagation equation resulting from the slowly varying polarization approximation applied to the nonlinear Maxwell wave equation~\cite{geissler1999light} in the case of three-wave mixing, the results of which are displayed in \fref{fig:GDD_DEP}(a). This method uses a DFT to calculate the nonlinear polarization in time. The peak intensity is kept constant at $I_0 = 1$\,GW/cm$^2$ with a crystal length of $200$\,$\mu$m, equivalent to that used in the experimental measurements and sufficiently low to prevent saturation or additional nonlinearities. The second method uses the SPA according to \fref{eq:SPA_FROG} for $\tau=0$ and is displayed in \fref{fig:GDD_DEP}(b). In both cases, the SHG spectrum has been normalized to unity peak intensity at each GDD value. The absolute intensity difference and the root-mean-square (RMS) error are plotted in \fref{fig:GDD_DEP}(c and d) respectively. 

The normalized SPA SHG intensity is independent of GDD because the approximation assumes the spectral intensity is mapped linearly to time, and thus the SHG spectrum is simply the square of the fundamental spectrum scaled by the refractive indices and SHG frequency. The fine structure (i.e. spectral hole) and finite bandwidth requires that the GDD must be sufficiently large so that the chirped pulse is significantly stretched to ensure only a single quasi-monochromatic frequency is present at any given group delay, according to the criterion in \fref{eq:SPA_CRITERION}
and marked by the dotted red line in \fref{fig:GDD_DEP}(d). For GDD magnitudes below this criterion, multiple frequencies interact, resulting in a structured SHG spectrum. For larger GDD magnitudes, the actual SHG spectrum asymptotically approaches that calculated using the SPA. Note that the criterion must use the smallest significant spectral feature, which is the width of the spectral hole in this scenario, although this only provides a ``rule of thumb''. For smooth spectra, the FWHM bandwidth will suffice. If the magnitude of the GDD is too large, then the pulse is stretched outside the temporal window given by the sampling rate, and is indicated by the dotted green line in \fref{fig:GDD_DEP}(d).

\begin{figure}[htbp]
\centering
\fbox{\includegraphics[width=\linewidth]{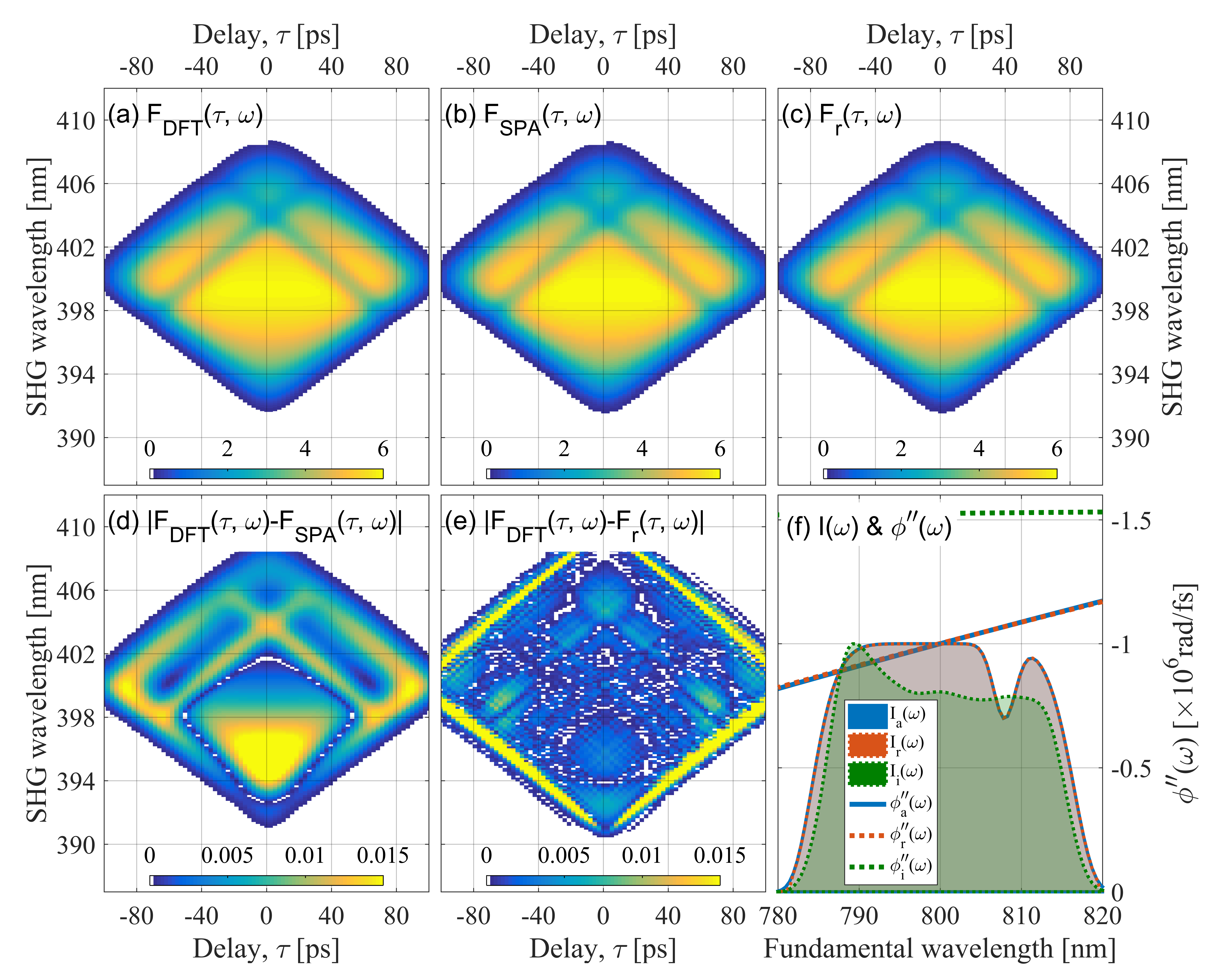}}
\caption{(a,b) Normalized SHG-FROG traces simulated using the DFT and SPA methods described for \fref{fig:GDD_DEP}. (c) Retrieved SHG-FROG trace. (d,e) Absolute intensity error between DFT FROG trace and the SPA/retrieved FROG traces respectively. (f) Spectral GDD (solid/dotted curves) and spectral intensity (shaded regions).}
\label{fig:SIM_FROG}
\end{figure}

\Fref{fig:SIM_FROG} shows the accuracy of using the SPA to calculate the SHG-FROG trace and the ability to reconstruct the fundamental pulse. The SHG spectrum as a function of delay between the two pulses was calculated using the same two procedures described above for \fref{fig:GDD_DEP} and using the same fundamental spectrum. The GDD was chosen to vary linearly as $\phi^{\prime\prime}(\omega) = 3\times10^6(\omega-\omega_0) - 10^6$\,fs$^2$/rad. The spectral intensity and spectral phase were retrieved using a downhill simplex minimization of \fref{eq:FROG_ERROR} using the FROG trace simulated via the DFT method as the ``measured'' trace. The main source of the error in the SPA simulation arises from the spectral hole and edges of the spectrum, where the spectral intensity varies rapidly. However, the error remains below $\sim 0.25$\% of the peak intensity. The error in the retrieved trace shows similar behavior in that the error is large where the spectral intensity varies rapidly, although the error at the edges of the trace are largest, most likely due to the low signal prevent an accurate retrieval. However, the maximum error is also $\sim 0.25$\% of the peak intensity. The FROG errors are 0.0021 for (d) and 0.0033 for (e).

The spectral intensity and GDD of the actual (blue), retrieved (red), and initial guess for the minimization routine (green) are plotted in \fref{fig:SIM_FROG}(f). The initial spectrum was calculated from the square root of the ``measured'' FROG trace integrated over all delays, and then divided by one plus the phase matching function (in order to minimize the effect of the nulls in the phase matching function). The initial GDD was calculated by the ratio of RMS width of the temporal marginal (measured FROG trace integrated over frequency) divided by the RMS width of the input spectral intensity. During minimization, the GDD was described by a linear polynomial and the spectral intensity was evenly sampled with 200 points.

\section{Experiment}
\label{sec:EXPERIMENT}

The method was tested experimentally by measuring the uncompressed output from a Ti:Sapphire CPA with a bandwidth of 55\,nm stretched to 32\,ps (both values corresponding to the full width at 1\% peak intensity), corresponding to a time-bandwidth product of $N\sim825$. This pulse, after frequency doubling in a 200\,\textmu{}m thick type I BBO crystal, is used as a pump in a CC-OPCPA~\cite{tang2008optical}. 
Accurate knowledge of the nonlinear chirp of this pulse is critical in the optimal design and implementation of the CC-OPCPA. In order to perform an SHG-FROG measurement, a Michelson interferometer was inserted into the beam path (before the lens that focuses the beam in to the BBO) to generate two spatially offset time-delayed replicas of the test pulse that spatially overlap when loosely focused into the crystal. After the crystal, the frequency mixing signal between the two input beams was spatially filtered and its spectrum measured on an single array spectrometer --- a detailed description can be found in \cite{delong1994frequency}.

\begin{figure}[htbp]
\centering
\fbox{\includegraphics[width=\linewidth]{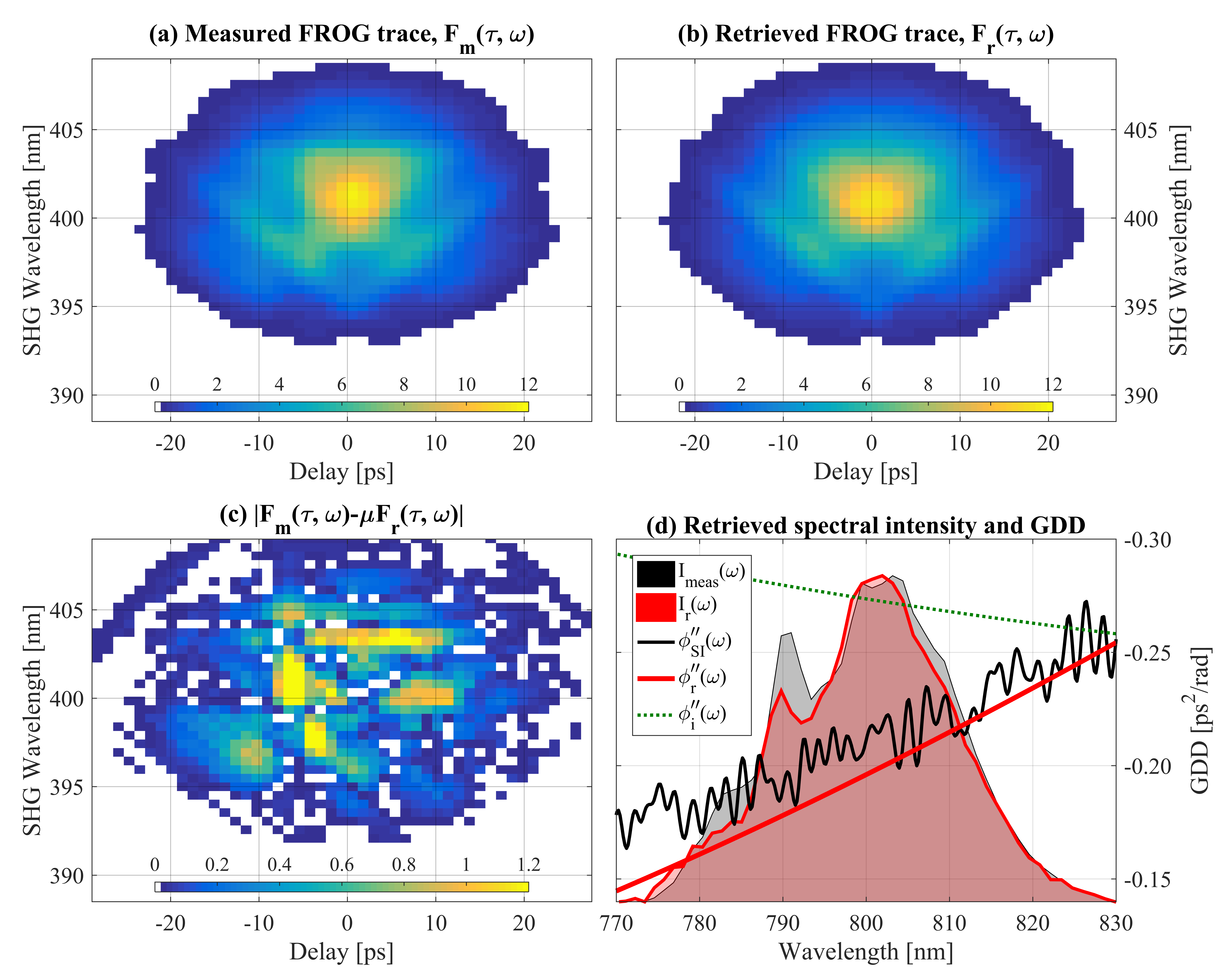}}
\caption{(a) Measured FROG trace after low-pass filtering and (b) retrieved FROG trace after optimizing for second order GDD, spectral intensity, zero delay position, crystal thickness and phase matching angle. (d) Absolute intensity error between normalized measured and retrieved FROG traces. (c) Measured (shaded gray) and retrieved (solid red) spectral intensity. Initial (dotted green curve), retrieved (solid red curve) and independently estimated (solid black curve) GDD.}
\label{fig:EXPT_RESULTS}
\end{figure}

The measured SHG-FROG trace is shown in \fref{fig:EXPT_RESULTS}(a) after some basic processing to remove noise (bandpass filtering followed by background subtraction and intensity calibration). \Fref{fig:EXPT_RESULTS}(b) shows the retrieved FROG trace using a downhill simplex minimization of \fref{eq:FROG_ERROR}. The agreement between the two traces is excellent, as indicated by the difference plot in \Fref{fig:EXPT_RESULTS}(c). The FROG error was 0.0996. The initial and retrieved GDD and spectral intensity are plotted in \fref{fig:EXPT_RESULTS}(d). Also plotted is an independent measurement of the fundamental spectrum and an independent estimate of the GDD measured using spectral interferometry (SI) between the input and output pulses of the stretcher used before the laser amplifier. Assuming near Fourier transform limited pulses entering the stretcher, an estimate of the GDD of the CPA was then added to the GDD of the stretcher as measured using SI. The offset between the two GDD curves is within the error of the estimate of the GDD using SI and dispersion calculations of the system.

The free parameters of the minimization routine were: (1) polynomial co-efficients of the GDD up to second order, (2) the fundamental spectral intensity values sampled evenly between 770--830\,nm with 60 points, (3) the phase matching angle (PMA), (4) the crystal thickness and (5) the zero delay offset. The minimisation routine requires an initial starting value for each free parameter. The zeroth order GDD coefficient was estimated by dividing the negative RMS temporal width of the spectrally integrated FROG trace by the RMS width of the measured spectrum, yielding a value of $\overline{\phi^{\prime\prime}} \sim -2.75$\,fs$^2$/rad.
During minimization, the magnitude of the zeroth order GDD coefficient was constraint to be larger than half of the initial value, and its sign fixed. Higher order GDD coefficients were set to $-2\times10^5$ with no constraint on their magnitude or sign. 
The measured spectral intensity was normalized to unity peak intensity and re-sampled as the initial spectrum. The initial crystal length and PMA was taken from the manufacturer's specification. The PMA could be further adjusted by comparing the measured SHG spectrum from one arm of the FROG interferometer with the square of the measured fundamental spectrum multiplied by a phase matching function. An alternative procedure was to measure the angle of the back-reflection from the crystal to provide the angle of incidence, and then calculating the angle of refraction of the central wavelength inside the crystal. Both of these two procedures gave very good agreement. The time-delay corresponding to the maximum measured SHG signal was taken as the zero time-delay.

\section{Conclusions}
\label{sec:CONCLUSIONS}

It has been shown that the SPA can be used to simplify the calculation of FROG measurements of highly chirped pulses. It was possible to accurately obtain higher order polynomial coefficients of the dispersion of the pulse, up to fourth order in the spectral phase, which could not have been obtained using the autocorrelation trace alone, which overestimated the average dispersion by $\sim 50$\%. The procedure also accurately retrieves the fundamental spectrum, thus also alleviating the requirement to measure this separately. These results have proven to be useful for us to model and optimise our CC-OPCPA system and it is expected it could be useful in other large bandwidth OPCPA systems.

\paragraph{Acknowledgements.} The authors would like to thank E. Springate for her comments on the paper.


\end{document}